\documentclass[conference]{IEEEtran}
\IEEEoverridecommandlockouts
\usepackage{amsmath,amssymb,amsfonts}
\usepackage{algorithmic}
\usepackage{graphicx}
\usepackage{subfigure}
\usepackage{subfig}
\usepackage{textcomp}
\usepackage{xcolor}
\usepackage{booktabs}
\usepackage{multirow}
\usepackage{floatrow}
\usepackage{url}
\newfloatcommand{capbtabbox}{table}[][\FBwidth]
\begin{document}

\title{Heterogeneous Graph Attention Networks for Early Detection of Rumors on Twitter}

\author{\IEEEauthorblockN{Qi Huang\IEEEauthorrefmark{1}\IEEEauthorrefmark{2},
		Junshuai Yu\IEEEauthorrefmark{1}\IEEEauthorrefmark{2},
		Jia Wu\IEEEauthorrefmark{3},
		and
		Bin Wang\IEEEauthorrefmark{5}
	}
	\IEEEauthorblockA{
\IEEEauthorrefmark{1}Institute of Information Engineering, Chinese Academy of Sciences, Beijing, China\\
\IEEEauthorrefmark{2}School of Cyber Security, University of Chinese Academy of Sciences, Beijing, China\\
\IEEEauthorrefmark{3}Department of Computing, Macquarie University, Sydney, Australia\\
\IEEEauthorrefmark{5}Xiaomi AI Lab, Beijing, China\\
\{huangqi, yujunshuai\}@iie.ac.cn, jia.wu@ieee.org, wangbin11@xiaomi.com
}
}

\maketitle

\begin{abstract}
With the rapid development of mobile Internet technology and the widespread use of mobile devices, it becomes much easier for people to express their opinions on social media. The openness and convenience of social media platforms provide a free expression for people but also cause new social problems. The widespread of false rumors on social media can bring about the panic of the public and damage personal reputation, which makes rumor automatic detection technology become particularly necessary. The majority of existing methods for rumor detection focus on mining effective features from text contents, user profiles, and patterns of propagation. 
Nevertheless, these methods do not take full advantage of global semantic relations of the text contents, which characterize the semantic commonality of rumors as a key factor for detecting rumors. In this paper, we construct a tweet-word-user heterogeneous graph based on the text contents and the source tweet propagations of rumors. A meta-path based heterogeneous graph attention network framework is proposed to capture the global semantic relations of text contents, together with the global structure information of source tweet propagations for rumor detection. Experiments on real-world Twitter data demonstrate the superiority of the proposed approach, which also has a comparable ability to detect rumors at a very early stage.
\end{abstract}


\section{Introduction}
With the rapid development of mobile Internet technology and the widespread use of mobile devices, it becomes much easier for people to express their opinions on social media. The openness and convenience of social media platforms provide a free expression for people but also result in new social problems caused by false rumor information on social media. A rumor is defined here as a statement or story whose true value is unverifiable or intentionally false at the time of publication \cite{difonzo2007rumor}. The widespread of rumors can bring about the panic of the public and denigrate an individual reputation. For example, on April 23, 2013 \footnote{\url{https://www.cnbc.com/id/100646197}}, a hacker hacked the Associated Press Twitter account and published a piece of breaking news saying that the White House had two explosions and Obama was injured. This news was widely spread on social media just in a few seconds, which caused extreme public panic and led to turbulence in the US stock market. Therefore, the automatic detection technology is desired for detecting rumors accurately and timely.

\begin{figure}[!t]
\centering
\includegraphics[scale=0.9]{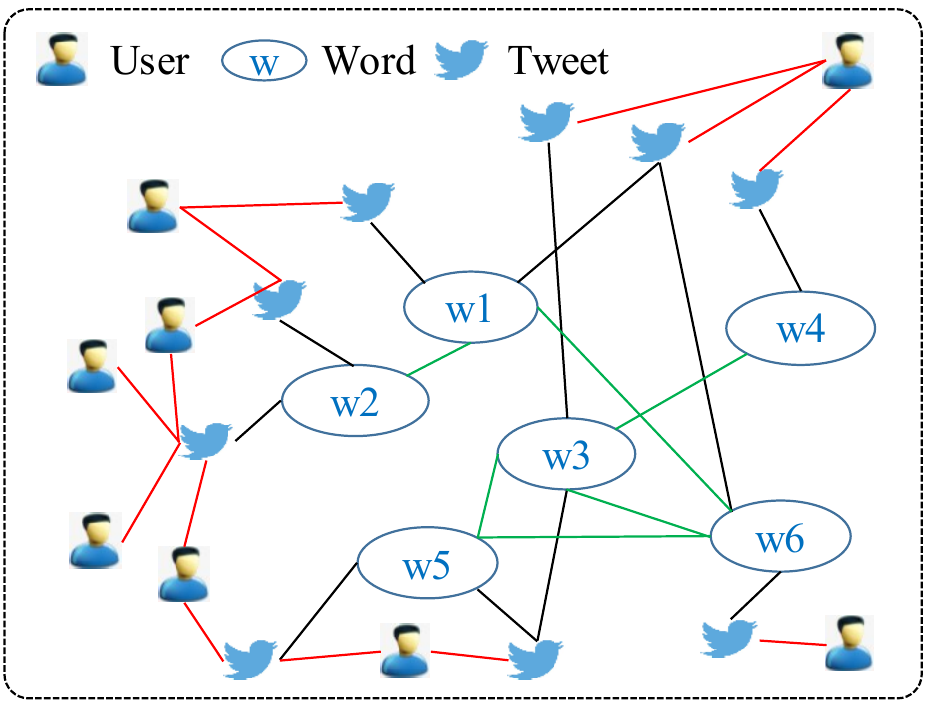}
\protect\caption{The heterogeneous tweet-word-user graph. Nodes involve words, tweets, and users. The edge (\textit{i.e.,} the green line) between two word nodes is built with the word co-occurrence information. The edge (\textit{i.e.,} the black line) between a word node and a tweet node is built with the word frequency and the tweet frequency of the word. And the edge (\textit{i.e.,} the red line) between a source tweet node and a user node is built with the behavior that the user retweeted or replied to tweets related to the source tweet. The detailed construction can be found in Section \ref{sec:HTG}.}
\label{Fig. 1}
\end{figure}

Most of the early detection methods for rumors utilized feature engineering to extract discriminated features from text contents \cite{castillo2011information,qazvinian2011rumor,popat2017assessing}, user profiles \cite{castillo2011information,yang2012automatic}, and propagation patterns \cite{jin2013epidemiological,sampson2016leveraging,ma2017detect}. Recently, inspired by the success of deep neural networks on feature exploration in many natural language processing tasks, such as sentiment analysis, machine translation, and text classification, Ma et al. \cite{ma2016detecting} exploited a recurrent neural network to capture the semantical variation of each source tweet and its retweets and make the prediction according to the semantical variation. This is the first work that introduces deep neural networks to capture the temporal representation of rumor source tweet propagations. Furthermore, Ma et al. \cite{ ma2018rumor} explored a tree-based recursive neural network to capture the semantic information and propagation clues of source tweet propagations for rumor detection. Yuan et al. \cite{yuan2019jointly} explored a global-local attention network to capture the local semantic relation and global structural information of the source tweet propagations for rumor detection. For these methods, one main limitation is that they neglect the global semantic relations of the text contents which however have been proved as useful \cite{yao2019graph}.

Indeed, social psychologists believe that rumors arise in the contexts of ambiguity when the situation is not obvious, or potential threat, when people need a sense of security. Therefore, the text contents of rumors tend to contain more ambiguous and intimidating words to promote the widespread of rumors. However, most of the previous rumor detection methods focus on the local semantic relations of the text contents in rumor propagations but ignore the global semantic relation between the text contents of different rumors.

In this paper, we present a novel meta-path based heterogeneous graph attention network framework to capture the global semantic relations of text contents and fuse the information involved in source tweet propagations for rumor detection. We first construct a heterogeneous tweet-word-user graph based on the text contents and the source tweet propagations of rumors, as shown in Fig. \ref{Fig. 1}. Then we decompose the heterogeneous graph into a tweet-word subgraph and a tweet-user subgraph based on the tweet-word and tweet-user path and exploit the subgraph attention network to learn the representation of nodes. Finally, we introduce an attention mechanism to fuse the representation of nodes in subgraphs for rumor detection.

The contributions of our paper are concluded in the following three aspects:
\begin{itemize}
\item This is the first work that constructs the text content and the source propagation of rumors as a heterogeneous tweet-word-user graph, where contains tweet, word, and user nodes, as shown in Fig. \ref{Fig. 1}.
\item We explore a novel meta-path based heterogeneous graph attention network framework to capture the global semantic relations of text contents and integrate them with the information involved in source tweet propagations for rumor detection.
\item The experiments on real-world Twitter datasets demonstrate that the proposed method outperforms the state-of-the-art baselines and has a comparable ability in detecting rumors at an early stage.
\end{itemize}

\section{Preliminaries}
In this section, we present the construction of a heterogeneous tweet-word-user graph and formulate the rumor detection problem on the heterogeneous graph.

\subsection{Heterogeneous Tweet-word-user Graph} \label{sec:HTG}

We build the rumor dataset as a heterogeneous tweet-word-user graph, which includes the text contents and the information involved in source tweet propagations of rumors. The structure of the heterogeneous graph is shown in Fig. \ref{Fig. 1}. For the constructed heterogeneous tweet-word-user graph $G=(V,E)$, where $V$ and $E$ denote the nodes and edges in the graph. The nodes $V$ consists of the set of source tweets $\mathcal{T}$, the set of words $\mathcal{W}$ that source tweets contained, and the set of users $\mathcal{U}$. The Edges $E$ have three types: tweet-word edges $E_{tw}$, word-word edges $E_{ww}$, and tweet-user edges $E_{tu}$. $E_{tw}$ describes the relationship that the tweet contains the word, $E_{ww}$ expresses the semantic relation between words, and $E_{tu}$ reflects the interaction between users and tweets.

\begin{figure*}[ht]
\centering
\includegraphics[scale=0.61]{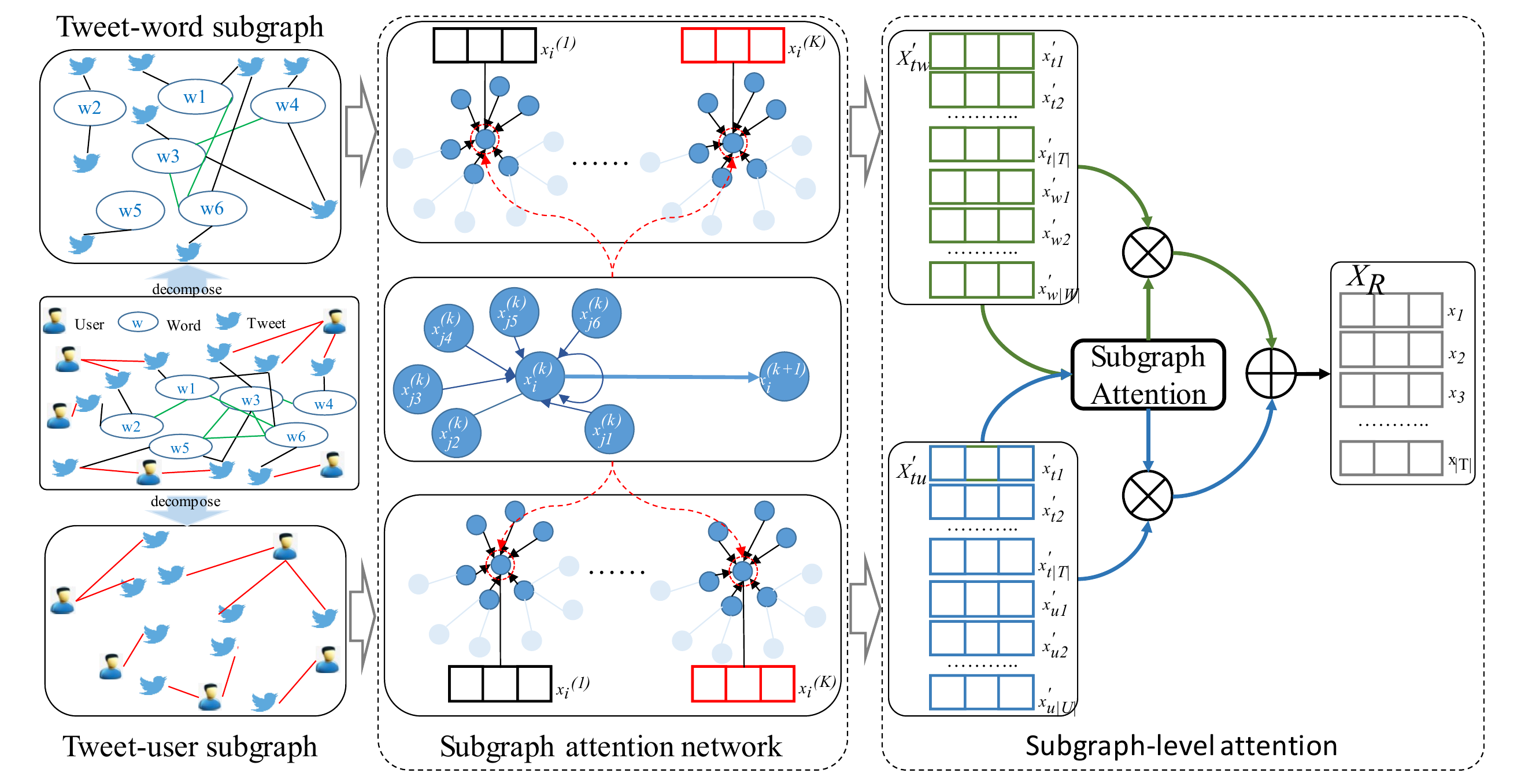}
\protect\caption{The architecture of our heterogeneous graph attention networks for rumor detection.}
\label{Fig. 2}
\end{figure*}

Specifically, We build the edges $E_{tw}$ with the word occurrence in source tweets, the edges $E_{ww}$ with the word co-occurrence information, and the edges $E_{tu}$ with the behavior that the user retweeted or replied to tweets related to the source tweet. The weight of the edges $E_{tw}$ is calculated from the term frequency-inverse document frequency (TF-IDF) of the word in the source tweet, where the term frequency is the number of times the word occurrences in the source tweet, and the inverse document frequency is the ratio of tweets total number and the number of tweets that contain the word. In order to use the global co-occurrence information of words, we set a fixed-size sliding window on all source tweets in the dataset to collect word co-occurrence statistics. And a popular measure for word associations, point-wise mutual information (PMI) \cite{zhang2019dual}, is exploited to calculate the weight of the edges $E_{ww}$. The weight of the edges $E_{tu}$ is the reciprocal of the time that the user retweeted or replied to tweets related to the source tweet. Formally, the weight of edges between node $i$ and node $j$ is calculated as follows:
\begin{equation}
A_{ij} =
	\begin{cases}
	\rm PMI(\mathit{i},\mathit{j}), &\rm \mathit{i},\,\mathit{j}\,are\,words,\,PMI(\mathit{i},\mathit{j})>0,\cr
	\rm TF{-}IDF_{\mathit{ij}},&\rm \mathit{i}\,is\,tweet,\,\mathit{j}\,is\,word,\cr
	\frac{1}{t+1},&\rm \mathit{i}\,is\,tweet,\mathit{j}\,is\,user,\cr
	1, &\rm \mathit{i}= \mathit{j},\cr
	0, &\rm otherwise.
	\end{cases}
\end{equation}
Where $t$ represents the elapsed time when a user $j$ retweeted or replied to tweets related to a source tweet $i$. The PMI value \cite{shi2019fine} of a word pair $\mathit{i}$, $\mathit{j}$ is calculated as 
\begin{equation}
\begin{aligned}
\rm PMI(\mathit{i},\mathit{j})=&log\frac{p(i,j)}{p(i)p(j)}\\
p(i,j)=&\frac{\#W(i,j)}{\#W}\\
p(i)=&\frac{\#W(i)}{\#W}
\end{aligned}
\end{equation}
where $\#W(i,j)$ denotes the number of sliding windows that contain both word $\mathit{i}$ and word $\mathit{j}$, $\#W$ represents the number of sliding windows, and $\#W(i)$ indicates the number of sliding windows that contain word $\mathit{i}$. And the TF-IDF value of a tweet $\mathit{i}$ and a word $\mathit{j}$ is calculated as follows:
\begin{equation}
\begin{aligned}
\rm TF{-}IDF_{\mathit{ij}}=&\rm TF_{\mathit{ij}}\times IDF_{\mathit{j}}\\
\rm TF_{\mathit{ij}}=&\frac{n_{ij}}{\sum_{k}n_{ik}}\\
\rm IDF_{\mathit{j}}=&\mathit{log}\frac{|\mathcal{T}|}{|\{k:w_j\in t_k\}|}
\end{aligned}
\end{equation}
where $n_{ij}$ represents the number of times that a word $\mathit{j}$ occurrences in a tweet $\mathit{i}$, $|\mathcal{T}|$ denotes the total number of tweets, $|\{k:w_j\in t_k\}|$ indicates the number of tweets that contain the word $\mathit{j}$.

\subsection{Subgraphs in Heterogeneous Tweet-word-user Graph}

To capture the global semantic relation of text contents and the information involved in source tweet propagations, we decompose the heterogeneous graph into a tweet-word subgraph and a tweet-user subgraph based on meta-path tweet-word and tweet-user.

\begin{itemize}
  \item \textbf{Tweet-word Subgraph:} the nodes in the tweet-word subgraph are the tweet and word nodes in the heterogeneous graph. The edges between tweets and words are consistent with the edges on the heterogeneous graph.
  \item \textbf{Tweet-user Subgraph:} the nodes in the tweet-user subgraph are the tweet and user nodes in the heterogeneous graph, and the edges are composed of the edges between tweets and users on the heterogeneous graph.
\end{itemize}

\subsection{Problem Statement}
Given a constructed heterogeneous tweet-word-user graph $G=(V,E)$, where $V=\{\mathcal{T},\mathcal{W},\mathcal{U}\}$ and $E=\{E_{tw},E_{ww},E_{tu}\}$ denotes nodes and edges in the graph. $\mathcal{T}$ denotes the set of source tweets, i.e., $\mathcal{T}=\{t_1,t_2,\cdots,t_n\}$, $\mathit{n}$ is the number of rumors. $\mathcal{W}$ represents the set of words contained in the source tweets, i.e., $\mathcal{W}=\{w_1,w_2,\cdots,w_{|\mathcal{W}|}\}$. And $\mathcal{U}$ indicates the set of social media users. $E_{tw},\,E_{ww},\,E_{tu}$ means the edges of tweet-word, word-word, and tweet-user respectively.

Our goal \textbf{aims} to learn a function $p(c|t_i,G;\theta)$ to determine the label probability of a source tweet $t_i$. $\mathit{c}$ and $\theta$ indicate the category label and the model parameters to be learned respectively.

\section{The Proposed Method}
We present a heterogeneous graph attention network framework to solve the problem of rumor detection on a heterogeneous graph. As shown in Fig. \ref{Fig. 2}, the framework includes a subgraph attention network and subgraph-level attention. The subgraph attention network exploits an attention mechanism similar to graph attention network \cite{velivckovic2017graph} for capturing the global relation information of nodes. The subgraph-level attention introduces an attention mechanism to fuse the source tweet representation in different subgraphs for rumor detection.

\subsection{Subgraph Attention Network}
Considering that the neighbors of each node in subgraphs have different importance to learn node embedding for rumor detection and inspired by graph attention networks \cite{velivckovic2017graph}, we present a subgraph attention network that utilizes an attention mechanism to learn the importance of each node's neighbors and merge the representation of these neighbors with the importance to form each node's representation.

In our constructed heterogeneous tweet-word-user graph, the set of words $\mathcal{W}$ denotes as $\mathcal{X}_{\mathcal{W}}=\{x_{w_1},x_{w_2},\cdots,x_{w_{|\mathcal{W}|}}\}, x_{w_i}\in\mathbb{R}^N$, $x_{w_i}$ is the word embedding of a word $w_i$, and $N$ is the dimension of word embedding. The set of tweets $\mathcal{T}$ represents as $\mathcal{X}_{\mathcal{T}}=\{x_{t_1},x_{t_2},\cdots,x_{t_{|\mathcal{T}|}}\},x_{t_i}\in\mathbb{R}^N$, where the representation $x_{t_i}$ of each tweet $t_i$ is calculated 
by the average of word representation contained, \textit{i.e.,} $x_{t_i} = \frac{1}{|t_i|}\sum_{w_j\in t_i}x_{w_j}$.  The set of users $\mathcal{U}$ indicates as $\mathcal{X}_{\mathcal{U}}=\{x_{u_1},x_{u_2},\cdots,x_{u_{|\mathcal{U}|}}\}, x_{u_i}\in\mathbb{R}^F$, the representation $x_{u_i}$ of each user $u_i$ can be extracted from the user behaviors or user profile data, and $F$ is the dimension of user features. If $u_i$ is not available, we initialize it by a normal distribution. Our representation method makes different types of nodes in the tweet-word subgraph have the same feature space and different types of nodes in the tweet-user subgraph have different feature space. Therefore, for the tweet and user nodes on the tweet-user subgraph, we design the transformation matrix $M_{\Phi_t}$ and $M_{\Phi_u}$ to project the representation of tweet and user nodes into the same vector space. The projection process is formalized as follows:

\begin{equation}
\mathcal{X}^{\prime}_{\mathcal{T}(\mathcal{U})}=M_{\Phi_{t(u)}}\cdot(\mathcal{X}_{\mathcal{T}(\mathcal{U})}+\mathcal{X}^0_{\mathcal{T}(\mathcal{U})})
\end{equation}
where $\mathcal{X}^0_{\mathcal{T}(\mathcal{U})}\in\mathbb{R}^{|\mathcal{T}|(|\mathcal{U}|)\times N(F)}$ is a dynamic vector and the value can be updated via the gradient, $\mathcal{X}^{\prime}_{\mathcal{T}(\mathcal{U})}$ and $\mathcal{X}_{\mathcal{T}(\mathcal{U})}$ denote the original and projected representation of the tweet (user) nodes on the tweet-user subgraph respectively. By using the transformation matrix, our subgraph attention network can handle the subgraph whose nodes have different feature spaces.

After that, the nodes in tweet-word subgraph denote as $\mathcal{X}_{tw}=\{x_{t_1},x_{t_2},\cdots,x_{t_{|\mathcal{T}|}},x_{w_1},x_{w_2},\cdots,x_{w_{|\mathcal{W}|}}\}, x_{t_i}\in\mathcal{X}_{\mathcal{T}}, x_{w_i}\in\mathcal{X}_{\mathcal{W}}$, and the nodes in tweet-user subgraph indicate as $\mathcal{X}_{tu}=\{x_{t_1},x_{t_2},\cdots,x_{t_{|\mathcal{T}|}},x_{u_1},x_{u_2},\cdots,x_{u_{|\mathcal{U}|}}\},x_{t_i}\in\mathcal{X}^{\prime}_{\mathcal{T}},x_{u_i}\in\mathcal{X}^{\prime}_{\mathcal{U}}$. Then we learn the weight among nodes in subgraphs utilizing a self-attention \cite{vaswani2017attention}. Given a node pairs $(i, j)$ in subgraphs, the self-attention mechanism $f$ can learn the attention coefficients $e_{i,j}$ which denotes how important the node $j$’ representation for node $i$. The attention coefficients $e_{i,j}$ of node pairs ($i$, $j$) can be calculated as follows:
\begin{equation}
e_{i,j}=f(Wx_i,Wx_j),\,x_i,x_j\in\mathcal{X}_{tw(tu)}
\label{eq.5}
\end{equation}
here $f$ can be implemented by a single-layer feedforward neural network that is parametrized by a weight vector \textbf{a} and applies the LeakyReLU \cite{xu2015empirical} as the activation function. $W$ denotes the weight matrix of a shared linear transformation.

Then, we introduce the structure information of the subgraph into the model by masked attention which indicates that we only compute the attention coefficients $e_{i,j}$ for nodes $j\in\mathcal{N}_i$, where $\mathcal{N}_i$ indicates the neighbors of the node $\mathit{i}$ in the subgraph (include itself). After getting the weight score between node pairs in the subgraph, we utilize the softmax function to normalize them for obtaining the coefficients $\alpha_{i,j}$:
\begin{equation}
\alpha_{i,j}=softmax(e_{i,j})=\frac{exp(\sigma(\textbf{a}^{T}\cdot[Wx_i||Wx_j]))}{\sum\limits_{k\in\mathcal{N}_i}exp(\sigma(\textbf{a}^T\cdot[Wx_i||Wx_k]))}
\end{equation}
where $\sigma$ means the activation function, and we apply LeakyReLU as the activation function in our experiments. $\textbf{a}$ denotes the weight vector of the neural network, $\cdot^{T}$ represents the transpose operation, $||$ indicates the concatenation operation.

Then, we aggregate the neighbor representations of a node $\mathit{i}$ in the subgraph and their corresponding coefficients to update the embedding of a node $\mathit{i}$ as follows:
\begin{equation}
x^{(1)}_i=\sigma(\sum\limits_{j\in\mathcal{N}_i}\alpha_{i,j}Wx_j)
\label{eq. 7}
\end{equation}
where $x^{(1)}_i$ denotes the updated embedding of a node $\mathit{i}$, $\sigma(\cdot)$ is a nonlinear function, $\mathit{W}$ represents the same weight matrix as Equation \ref{eq.5}, and $\mathcal{N}_i$ indicates the set which contains a node $\mathit{i}$ and its neighbors.

Finally, we extend employing a self-attention to multi-head attention similar to graph attention networks \cite{velivckovic2017graph} for learning a more stable embedding. Specifically, we execute the transformation of Equation \ref{eq. 7} for $\mathit{K}$ times and concatenate their learned representation to obtain the final output representation:
\begin{equation}
x^{\prime}_i=\overset{K}{\underset{k=1}{||}}\sigma(\sum\limits_{j\in\mathcal{N}_i}\alpha_{i,j}^kW^kx_j)
\end{equation}
where $||$ means the concatenation operation, $\alpha_{i,j}^{k}$ denotes the normalized attention coefficients obtained by the $k$-th attention mechanism ($f^k$), and $W^k$ represents the corresponding weight matrix of input linear transformation.

Given the representation $\mathcal{X}_{tw}$ of nodes in tweet-word subgraph and the representation $\mathcal{X}_{tu}$ of nodes in tweet-user subgraph, after feeding the representation of nodes into subgraph attention neural network, we can obtain the node embeddings $\mathcal{X}^{\prime}_{tw}=\{x^{\prime}_{t_1},x^{\prime}_{t_2},\cdots,x^{\prime}_{t_{|\mathcal{T}|}},x^{\prime}_{w_1},x^{\prime}_{w_2},\cdots,x^{\prime}_{w_{|\mathcal{W}|}}\}$ of tweet-word subgraph and the node embeddings $\mathcal{X}^{\prime}_{tu}=\{x^{\prime}_{t_1},x^{\prime}_{t_2},\cdots,x^{\prime}_{t_{|\mathcal{T}|}},x^{\prime}_{u_1},x^{\prime}_{u_2},\cdots,x^{\prime}_{u_{|\mathcal{U}|}}\}$ of tweet-user subgraph with global relation information.

\subsection{Subgraph-level Attention}
The subgraphs decomposed according to the meta-path in the heterogeneous graph contain different information. The tweet-word subgraph contains the global semantic relation information of text contents, while the tweet-user subgraph contains the information involved in source tweet propagations. To accurately identify rumors, we need to fuse the information contained in two subgraphs. To this end, we present a novel subgraph-level attention mechanism to learn subgraph weights for rumor detection. Given the node embeddings $\mathcal{X}^{\prime}_{tw}$ and $\mathcal{X}^{\prime}_{tu}$ as input, the weights of the tweet-word and tweet-user subgraph are calculated as follows:
\begin{equation}
(\beta_{tw},\beta_{tu})=att_{sub}(\mathcal{X}^{\prime}_{tw},\mathcal{X}^{\prime}_{tu})
\end{equation}
where $\mathit{att_{sub}}$ represents a feedforward neural network that performs subgraph-level attention. 

In order to learn the weights of the tweet-word subgraph and the tweet-user subgraph, we first transform the representation of the node in subgraphs by a nonlinear transformation (e.g. single-layer MLP). Then we calculate the similarity between the transformed node representations and the subgraph-level attention vector $\textbf{a}$ as the importance of nodes. Furthermore, we average the importance of all nodes in subgraphs as the importance of subgraphs. The formula for calculating the importance $w_{tw(tu)}$ of tweet-word (tweet-user) subgraph is as follows:
\begin{equation}
w_{tw(tu)}=\frac{1}{|\mathcal{X}^{\prime}_{tw(tu)}|}\sum\limits_{x_i\in\mathcal{X}^{\prime}_{tw(tu)}}\textbf{a}^{T}\cdot tanh(W_{sub} x_i)
\end{equation}
where $W_{sub}$ represents the weight matrix, together with the subgraph-level attention vector $\textbf{a}$, are shared by the tweet-word subgraph and the tweet-user subgraph. After obtaining the importance of subgraphs, we exploit the softmax function to normalize the importance of subgraphs. The weight of tweet-word (tweet-user) subgraph, denoted as $\beta_{tw(tu)}$, can be calculated by normalizing the weights of two subgraphs using softmax function:
\begin{equation}
\beta_{tw(tu)}=\frac{exp(w_{tw(tu)})}{\sum\limits_{\Phi\in\{tw,tu\}}exp(w_{\Phi})}
\end{equation}

Finally, with the learned weight coefficients of subgraphs, we fuse the representation of tweet nodes in subgraphs to obtain the representation $\mathcal{X}_{\mathcal{T}}$ of source tweets as follows:
\begin{equation}
\begin{aligned}
\mathcal{X}_{\mathcal{T}}=&\{x_1,x_2,\cdots,x_{|\mathcal{T}|}\}\\
x_i=&\sum_{\Phi\in{tw,tu}}\beta_{\Phi}\cdot x_{t_i}, x_{t_i}\in\mathcal{X}^{\prime}_{\Phi}
\end{aligned}
\end{equation}
where $|\mathcal{T}|$ represents the total number of source tweets, $x_{t_i}$ means the representation of a tweet node $\mathit{i}$ with global relation information in the $\Phi$ subgraph, and $\mathcal{X}^{\prime}_{\Phi}$ denotes the representation of nodes with global relation information in the $\Phi$ subgraph.
\subsection{Rumor Detection}
In this work, we feed the representation $\mathcal{X}_{\mathcal{T}}$ of source tweets into a one-layer feedforward neural network (FNN) with softmax normalization, and thus predict the class probability distribution of source tweets by the following formula:
\begin{equation}
p(c|t_i,G;\theta)=\rm softmax(FNN(\mathit{x_i})),\mathit{x_i}\in\mathcal{X}_{\mathit{R}}
\end{equation}

To train the model's parameters, the cross-entropy loss and a regularization term are used as the model's objective optimization function. The function is formalized as follows:
\begin{equation}
\mathcal{L}=-\sum\limits_{i\in|\mathcal{T}|}y_i p(c|t_i,G;\theta)+\lambda\lVert\theta\rVert^2_2
\end{equation}
where $y_i$ denotes the ground truth one-hot vector of the $\mathit{i}$-th source tweet, $\lambda$ represents the trade-off coefficient and $\lVert\cdot\rVert^2_2$ indicates the L2 regularization term to prevent overfitting.

\section{Experiments}
In this section, we conduct comprehensive experiments to verify the performance of our proposed framework on (early) rumor detection tasks\footnote{The source codes and datasets used are available in the Github repository \url{https://github.com/201518018629031/HGATRD}.}.

\subsection{Datasets}
Experiments are based on two public available Twitter datasets. These datasets were collected by Ma et al. \cite{ma2017detect}, recorded as Twitter15 and Twitter16. Twitter15 and Twitter16 contained 1,490 and 818 source tweets of rumors, respectively (the more details are shown in Table \uppercase\expandafter{\romannumeral1}). Each source tweet in datasets is labeled as non-rumor, false rumor, true rumor, or unverified rumor \cite{zubiaga2016analysing}. To ensure the fairness of the evaluation, we follow the same setting in the previous works \cite{liu2018early,yuan2019jointly} that randomly select 10\% of datasets as the validation set for model selection, and the remaining is divided into the training and the testing set according to a ratio of 3:1. Since the original datasets do not include the user profile information, we call Twitter API\footnote{\url{https://dev.twitter.com/rest/public}} to crawl the profiles of all users related to source tweets.

\subsection{Experimental Settings}
We compare our proposed framework with the following ten rumor detection baselines:
\begin{itemize}
\item \textbf{DTR}: A Decision Tree-based model to identify rumors by ranking the clusters extracted from the Twitter stream via regular expressions \cite{zhao2015enquiring}.
\item \textbf{DTC}: A Decision-Tree classifier that exploits the statistical features of tweets extracted by feature engineering \cite{castillo2011information}.
\item \textbf{RFC}: A Random Forest classifier that exploits 3 fitting parameters as temporal attributes and a set of manual features on users, contents, and structural attributes \cite{kwon2013prominent}.
\item \textbf{SVM-TS}: A linear SVM classifier that utilizes the variation of manual social context features in Time Series \cite{ma2015detect}.
\item \textbf{SVM-HK}: An SVM classifier with hybrid kernel consisting of a random-walk-based graph kernel and an RBF kernel \cite{wu2015false}.
\item \textbf{SVM-TK}: An SVM classifier which exploits a tree-based kernel to compute the similarity of the propagation tree structure to identify rumors \cite{ma2017detect}.
\item \textbf{GRU-RNN}: RNN with GRU units was utilized to model the sequential structure of relevant tweets for rumor detection \cite{ma2016detecting}.
\item \textbf{BU-RvNN and TD-RvNN}: A Recursive Neural Network based on the traversal direction of the propagation tree to capture propagation clues and content semantics \cite{ma2018rumor}.
\item \textbf{PPC}: A Propagation Path Classifier consisting of recurrent and convolutional networks to model the sequence of user characteristics \cite{liu2018early}.
\item \textbf{GLAN}: A Global-Local Attention Network to capture the local semantic relation and global structure information of the tweet propagation \cite{yuan2019jointly}.
\end{itemize}

\begin{table}[!t]
	\caption{Statistics of the datasets}
	\begin{center}
		\begin{tabular}{ccc}
			\toprule
			\textbf{Statistic}&\textbf{Twitter15}&\textbf{Twitter16}\\
			\midrule
			\# of source tweets &1,490&818\\
			\# of users&276,663&173,487\\
			\# of tweets&331,612&204,820\\
			\# of non-rumors&374&205\\
			\# of false-rumors&370&205\\
			\# of true-rumors&372&207\\
			\# of unverified rumors&374&201\\
			\bottomrule
		\end{tabular}
		\label{tab1}
	\end{center}
\end{table}

For a fair comparison, we utilize the micro-average accuracy (i.e., Acc.) in all categories and the $F_1$ score of the precision and recall in each category to evaluate the performance of models. Our framework is implemented by PyTorch. The parameters are updated using Adam algorithm \cite{kingma2014adam}. The learning rate is initialized at 0.005 and gradually decreases during the model training process. We select the best parameter settings based on the performance on the verification set and evaluate our framework performance on the test set. We initialize the word vector with 300 dimensions word embedding. The output dimension of the subgraph attention network is set to 300. The number of heads $K$ of the subgraph attention network is set to 8. The batch size is set to 64 during training.
\subsection{Rumor Detection}
As shown in Table \uppercase\expandafter{\romannumeral2}, our method performs better than all other baselines on two datasets. Specifically, our proposed framework achieves an accuracy of 91.1\% and 92.4\% respectively, increasing by 2.1\% and 2.2\% compared with the best baseline. This demonstrates that our proposed framework can effectively capture the global semantic relations of the text contents in rumors, which is helpful for rumor detection. 

It is observed that the baselines based on traditional machine learning methods (\textit{i.e.,} DTR, DTC, RFC, SVM-TS, SVM-HK, and SVM-TK) do not perform well. Among these baseline methods, RFC, SVM-TS, SVM-HK, and SVM-TK are relatively better than DTR and DTC. That is mainly because they exploit additional timing or structural features.

For deep learning methods (\textit{e.g.,} BU-RvNN, TD-RvNN, PPC, and GLAN), all of them have better performance than the methods based on traditional machine learning, which shows that deep learning methods are easier to capture efficient features for rumor detection. We also observe that GLAN achieves the best performance among the baselines, for the reason that it captures the local semantics and global structure information of the source tweet propagation of rumors while others capture the part of this information.

\begin{table}
\caption{Experimental results of rumor detection. (NR, FR, TR, and UR denote non-rumor, false rumor, true rumor, and unverified rumor, respectively.)}\vspace{-1mm}
\begin{center}
\centerline{(a) Twitter15 dataset}
\vspace{1.5mm}
\begin{tabular}[c]{c|c|c|c|c|c}
\hline\hline
\multirow{2}{*}{Method}&\multirow{2}{*}{Acc.}&NR&FR&TR&UR\\
\cline{3-6}
&&$F_1$&$F_1$&$F_1$&$F_1$ \\
\hline \hline
DTR&0.409&0.501&0.311&0.364&0.473\\
DTC&0.454&0.733&0.355&0.317&0.415\\
RFC&0.565&0.810&0.422&0.401&0.543\\
SVM-TS&0.544&0.796&0.472&0.404&0.483\\
SVM-HK&0.493&0.650&0.439&0.342&0.336\\
SVM-TK&0.667&0.619&0.669&0.772&0.645\\
\hline
GRU-RNN&0.641&0.684&0.634&0.688&0.571\\
BU-RvNN&0.708&0.695&0.728&0.759&0.653\\
TD-RvNN&0.723&0.682&0.758&0.821&0.654\\
PPC&0.842&0.818&0.875&0.811&0.790\\
GLAN&0.890&0.936&0.908&0.897&0.817\\
\hline
Our method &\textbf{0.911}&\textbf{0.953}&\textbf{0.929}&\textbf{0.905}&\textbf{0.854}\\
\hline\hline
\end{tabular}

\vspace{3mm}
\centerline{(b) Twitter16 dataset}
\vspace{1.5mm}
\begin{tabular}[c]{c|c|c|c|c|c}
\hline\hline
\multirow{2}{*}{Method}&\multirow{2}{*}{Acc.}&NR&FR&TR&UR\\
\cline{3-6}
&&$F_1$&$F_1$&$F_1$&$F_1$ \\
\hline \hline
DTR&0.414&0.394&0.273&0.630&0.344\\
DTC&0.465&0.643&0.393&0.419&0.403\\
RFC&0.585&0.752&0.415&0.547&0.563\\
SVM-TS&0.574&0.755&0.420&0.571&0.526\\
SVM-HK&0.511&0.648&0.434&0.473&0.451\\
SVM-TK&0.662&0.643&0.623&0.783&0.655\\
\hline
GRU-RNN&0.633&0.617&0.715&0.577&0.527\\
BU-RvNN&0.718&0.723&0.712&0.779&0.659\\
TD-RvNN&0.737&0.662&0.743&0.835&0.708\\
PPC &0.863&0.843&0.898&0.820&0.837\\
GLAN&0.902&0.921&0.869&0.847&\textbf{0.968}\\
\hline
Our method&\textbf{0.924}&\textbf{0.935}&\textbf{0.913}&\textbf{0.947}&0.899\\
\hline\hline
\end{tabular}
\label{tab2}
\end{center}\vspace{-3mm}
\end{table}

\subsection{Early Rumor Detection}
The time for debunking rumors depends on the time when the rumors are detected, so the early detection of rumors is particularly important. To simulate the process of early rumor detection, we control the elapsed time or user-retweet count after the source tweet of the rumor was posted to represent the different periods of rumor propagation, and then calculate the accuracy of rumor detection at different periods to evaluate the performance. For the early detection of rumors, we compare our framework with five state-of-the-art baselines, \textit{i.e.,} RFC, SVM-TK, TD-RvNN, PPC, GLAN. 

\begin{figure*}[ht]
	\centering
	\subfigure[Twitter15 (elapsed time)]{
		\begin{minipage}[b]{0.23175\textwidth}
			\includegraphics[width=1\textwidth]{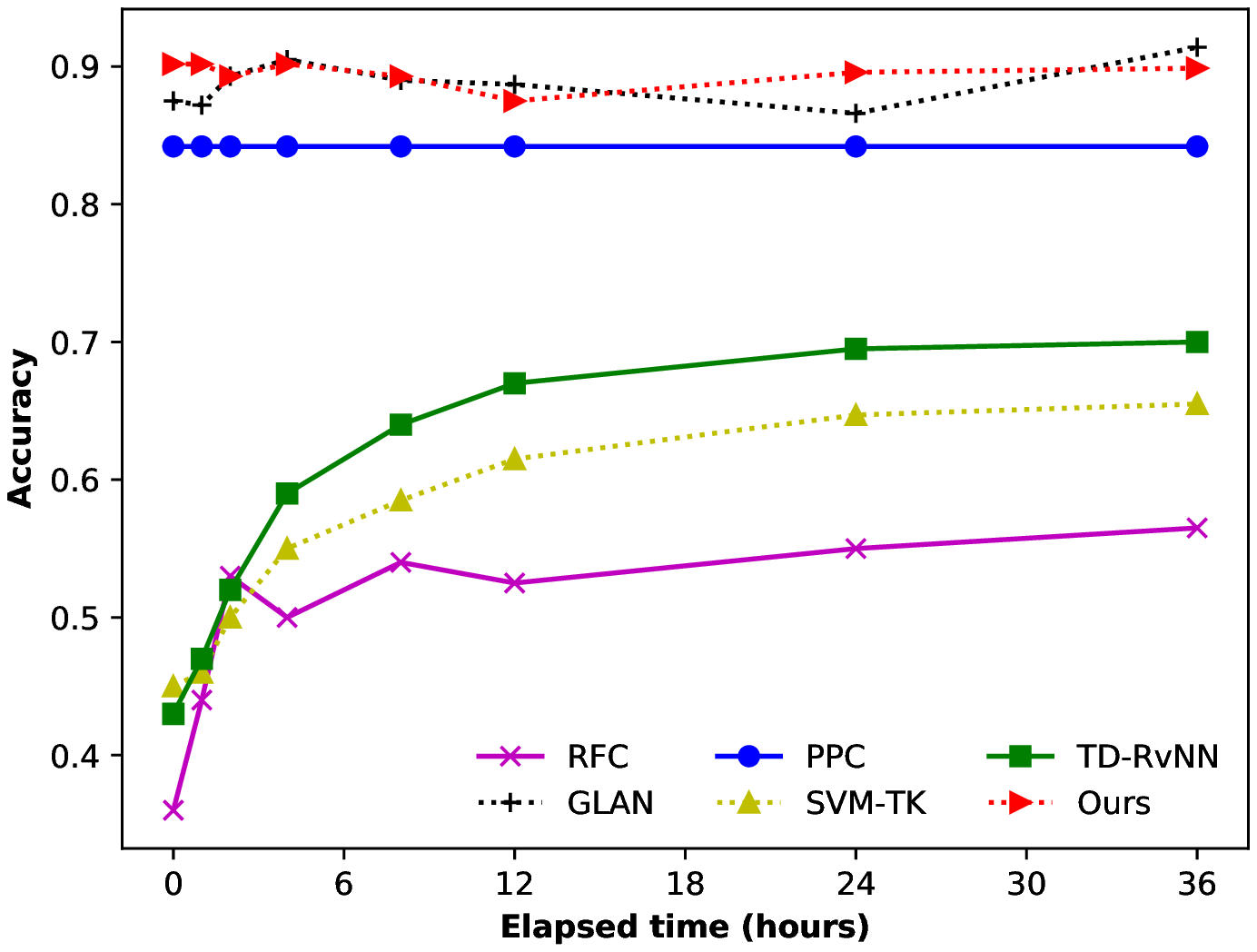} 
		\end{minipage}
	}
    	\subfigure[Twitter16 (elapsed time)]{
    		\begin{minipage}[b]{0.23175\textwidth}
   		 	\includegraphics[width=1\textwidth]{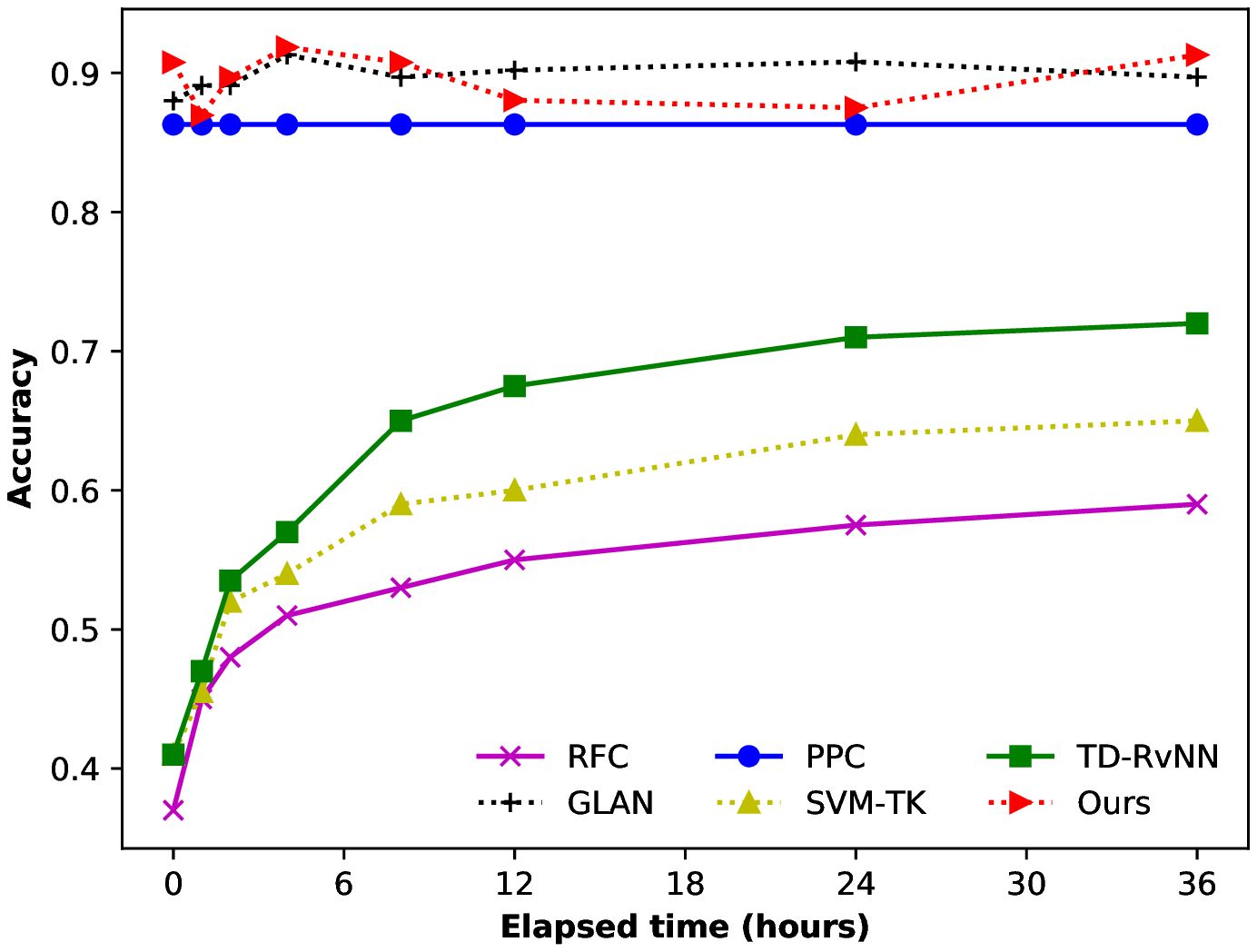}
    		\end{minipage}
    	}
	\subfigure[Twitter15 (user-retweet count)]{
		\begin{minipage}[b]{0.23175\textwidth}
			\includegraphics[width=1\textwidth]{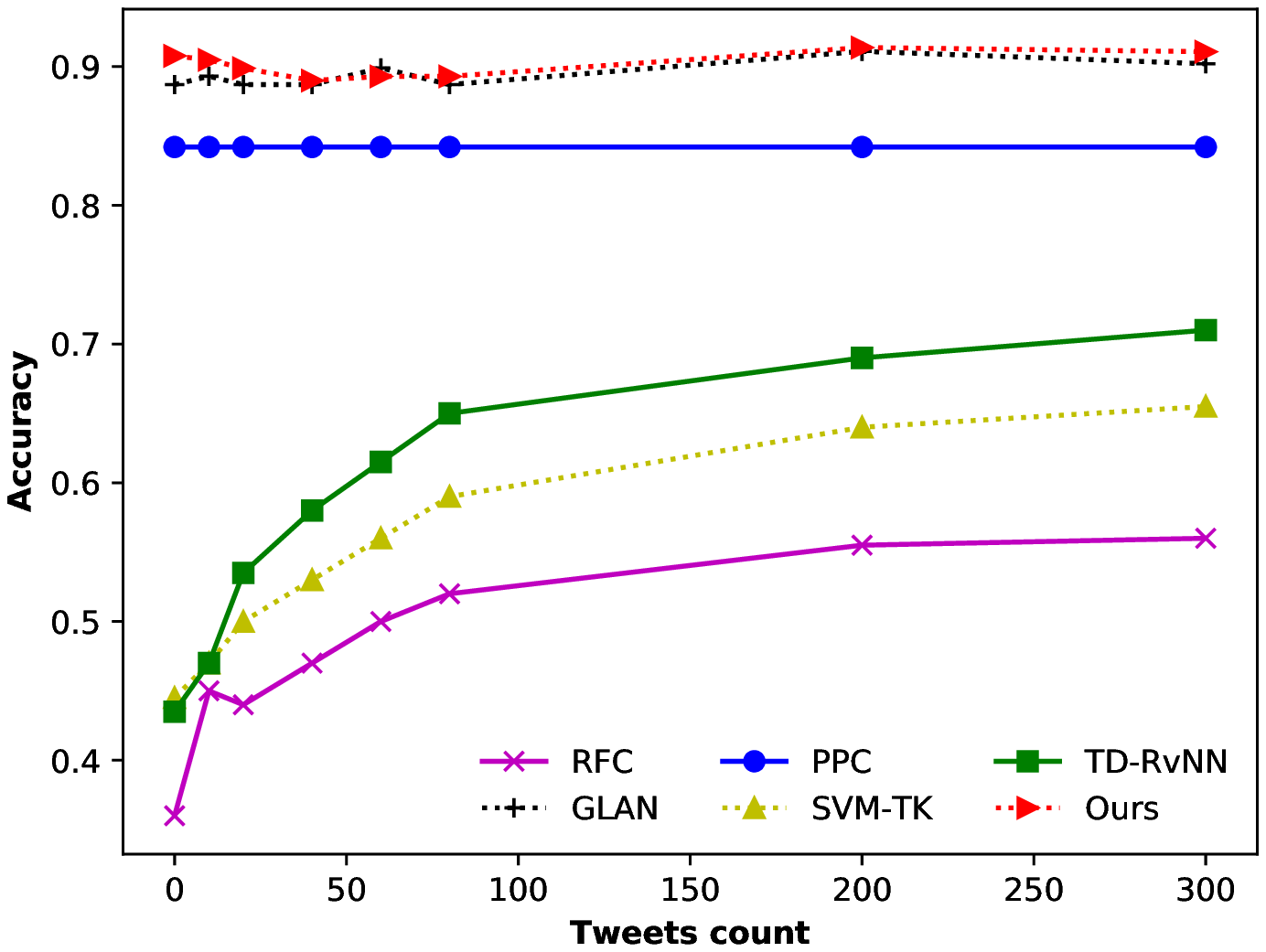} 
		\end{minipage}
	}
    	\subfigure[Twitter16 (user-retweet count)]{
    		\begin{minipage}[b]{0.23175\textwidth}
		 	\includegraphics[width=1\textwidth]{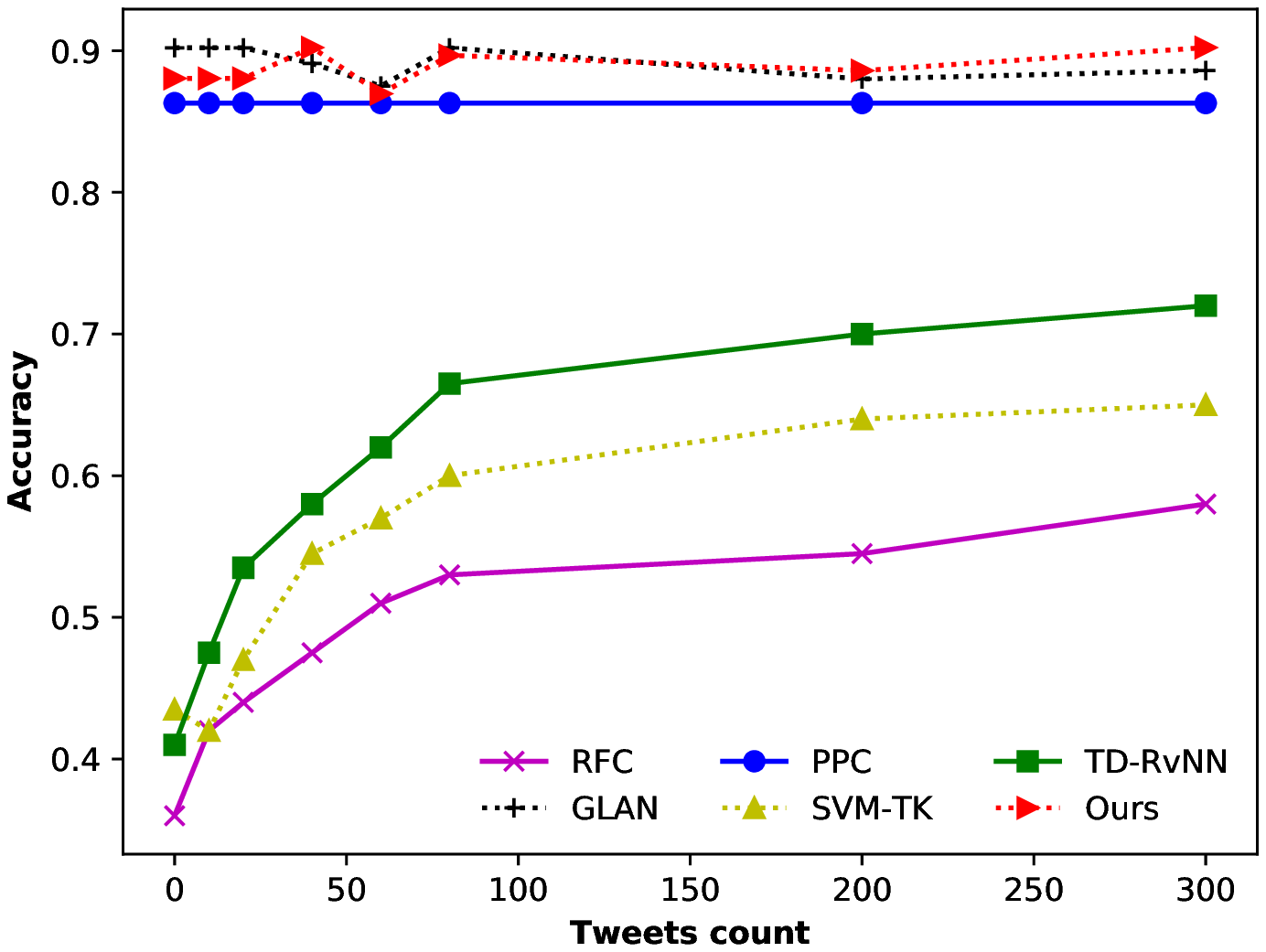}
    		\end{minipage}
    	}
	\caption{Accuracy of early rumor detection at different checkpoints in times of elapsed time (user-retweet count)}
	\label{fig:fig3}
\end{figure*}

Fig. \ref{fig:fig3} shows the accuracy of several competitive models in the various periods of rumor propagation simulated by the elapsed time or user-retweet count. Our method performs very well on early rumor detection task. Only though less than 1-hour [see Figs. \ref{fig:fig3}(a) and \ref{fig:fig3}(b)]  or 10-retweets [see Figs. \ref{fig:fig3}(c) and \ref{fig:fig3}(d)], our method has already outperformed the PPC using all data. This indicates that our method has a comparable ability to enable early detection of rumors. With the variation of the elapsed time or user-retweet count, our method and GLAN have a slight fluctuation in the accuracy of rumor detection. This is because as the information increases, it also brings noise information to the models. 

For an early detection task, a good solution should achieve an appropriate performance as early as possible. Based on that, our method performs better than GLAN in most cases. Therefore, the results show that our method is more robust to data and has a more stable performance.

To sum up, experimental results on two real-world Twitter datasets show that our method has better performance than the state-of-the-art baselines on (early) rumor detection.

\subsection{Importance Analysis of Subgraphs}
In order to analyze the importance of subgraphs decomposed from the heterogeneous tweet-word-user graph by meta-path for rumor detection, we model subgraphs exploiting the subgraph attention network and learn the node representations of subgraphs for rumor detection. The experimental results are shown in Table \uppercase\expandafter{\romannumeral3}.

\begin{itemize}
\item \textbf{Only tweet-word}: Modeling the tweet-word subgraph utilizing the subgraph attention network and learn tweet representations for rumor detection.
\item \textbf{Only tweet-user}: Modeling the tweet-user subgraph via the subgraph attention network and learn tweet representations for rumor detection.
\end{itemize}

\begin{table}
\caption{The importance analysis of subgraphs (NR, FR, TR, and UR denote non-rumor, false rumor, true rumor, and unverified rumor, respectively.)}
\begin{center}
\centerline{(a) Twitter15 dataset}
\begin{tabular}[c]{c|c|c|c|c|c}
\hline\hline
\multirow{2}{*}{Method}&\multirow{2}{*}{Acc.}&NR&FR&TR&UR\\
\cline{3-6}
&&$F_1$&$F_1$&$F_1$&$F_1$ \\
\hline \hline
All&0.911&0.953&0.929&0.905&0.854\\
Only tweet-word&0.813&0.708&0.857&0.864&0.813\\
Only tweet-user&0.560&0.812&0.430&0.544&0.446\\
\hline\hline
\end{tabular}

\vspace{3mm}
\centerline{(b) Twitter16 dataset}
\vspace{1mm}
\begin{tabular}[c]{c|c|c|c|c|c}
\hline\hline
\multirow{2}{*}{Method}&\multirow{2}{*}{Acc.}&NR&FR&TR&UR\\
\cline{3-6}
&&$F_1$&$F_1$&$F_1$&$F_1$ \\
\hline \hline
All&0.924&0.935&0.913&0.947&0.899\\
Only tweet-word&0.810&0.698&0.787&0.936&0.809\\
Only tweet-user&0.690&0.772&0.651&0.697&0.632\\
\hline\hline
\end{tabular}
\label{tab2}
\end{center}
\end{table}

From the experimental results in Table \uppercase\expandafter{\romannumeral3}, we can observe that:
\begin{itemize}
\item On the whole, for the constructed heterogeneous graph, the decomposed tweet-word subgraph including rumor text contents has a greater effect for rumor detection than the decomposed tweet-user subgraph which includes the source tweet propagation of rumors. Specifically, we can see that the accuracy of modeling tweet-word subgraph drops 9.8\% and 11.4\% while the accuracy of modeling tweet-user subgraph drops 35.1\% and 23.4\% on Twtter15 and Twitter16 datasets. This result indicates that the text content of rumors in our constructed heterogeneous graph shows more important for rumor detection.

\item Only for the non-rumor class, the decomposed tweet-user subgraph has a greater effect than the decomposed tweet-word subgraph. Specifically, the $F_1$ value of identifying non-rumor class by modeling tweet-user subgraph drops 14.1\% and 16.3\% while the $F_1$ value of identifying non-rumor class by modeling tweet-word subgraph drops 24.5\% and 23.7\% on Twtter15 and Twitter16 datasets. This is due to the fact that non-rumors are more likely to be responded by users who have higher credibility and their propagation paths are relatively fixed so it is easier to distinguish non-rumors from rumors by the structure of the source tweet propagation.
\end{itemize}

\section{Related Works}
\subsection{Traditional Machine Learning Methods}
The majority of early detection methods for rumors were based on statistical machine learning, which manually extracts effective features to identify rumor from the text contents, user profiles, and propagation patterns of the rumors. One class of these approaches combined different types \cite{zhao2015enquiring} or different types of temporal variations \cite{kwon2013prominent} of features to detect rumors. Castillo et al. \cite{castillo2011information} proposed a series of features consists of message-based, user-based, topic-based, and propagation-based for rumor detection, and Ma et al. \cite{ma2015detect} explored a novel approach to capture the temporal characteristics of social context features based on the time series of rumor's lifecycle. Another class exploited the topological-structure \cite{wu2015false} features extracted from the source tweet propagation of rumors to detect rumors. Ma et al. \cite{ma2017detect} presented an SVM classifier with a tree-based kernel function calculating the similarity of the propagation topological tree to identify rumors.

However, these methods are time-consuming and labor-intensive due to the features manually extracting from text contents, user profiles, and propagation patterns. And these features depend on the datasets and sometimes are impossible to be extracted.

\subsection{Deep Learning Methods}
Deep learning has been successfully applied in many real-world tasks, such as natural language processing. Researchers also begin to work on the deep neural networks for detecting rumors. One group of researchers exploited neural networks to capture the temporal information \cite{ma2016detecting} or topological-structure \cite{ma2018rumor} information of the source tweet propagation of rumors for rumor detection. Liu et al. \cite{liu2018early} modeled the source tweet propagation as a sequence of user characteristics and proposed a propagation path classifier composed of recurrence and convolutional networks to detect rumors. Yuan et al. \cite{yuan2019jointly} posed a global-local attention network (GLAN) to capture the global structure features of the source tweet propagation topological tree for rumor detection. Another group of researchers \cite{ruchansky2017csi} explored a framework consists of three modules to capture the information from text contents, user profiles, and propagation patterns for rumor detection. Huang et al. \cite{huang2019deep} utilized a graph convolutional network to model the user graph formed by user behaviors and combined the powerful user representation with the representation of the propagation tree for rumor detection on twitter.

However, these methods ignore the global semantic relation in the text content of rumors, and its integration with the information involved in the source tweet propagation effectively not been solved. In this paper, we construct a heterogeneous tweet-word-user graph according to the text content and the source tweet propagation of rumors and propose a new meta-path based heterogeneous graph attention network framework to capture the global semantic relation in text contents and effectively integrate it with the information involved in source tweet propagations for rumor detection.

\section{Conclusions}
Semantic relations among the text contents of rumors are ignored by the majority of existing rumor detection methods. To address this challenge, we constructed a heterogeneous tweet-word-user graph based on the text contents and the source tweet propagations of rumors. A meta-path based heterogeneous graph attention network framework was further proposed to learn the global semantic relations of text contents and effectively integrate them with the information related to source tweet propagations for rumor detection. 

Specifically, we first decomposed the heterogeneous graph into a tweet-word subgraph and a tweet-user subgraph according to the tweet-word and tweet-user meta path. Then the subgraph attention network was exploited to model the subgraphs for obtaining the representation of nodes with global structure information. Finally, we utilized an attention mechanism to integrate the representations of tweet nodes in different subgraphs for rumor detection. Experiments on two real-world Twitter datasets demonstrated that our method has better performance than the state-of-the-art baselines in accuracy and has a comparable ability on the early rumor detection task. 

As for the future works, how to integrate the social relationship of users into the heterogeneous tweet-word-user graph to improve the performance of early rumor detection is an emerging research problem.

\section*{Acknowledgment}
This work was partially supported by the National Key Research and Development Program of China (No. 2016YFB0801003) and the ARC DECRA Project (No. DE200100964).

\bibliographystyle{IEEEtran}
\bibliography{111}

\begin{thebibliography}{10}
\providecommand{\url}[1]{#1}
\csname url@samestyle\endcsname
\providecommand{\newblock}{\relax}
\providecommand{\bibinfo}[2]{#2}
\providecommand{\BIBentrySTDinterwordspacing}{\spaceskip=0pt\relax}
\providecommand{\BIBentryALTinterwordstretchfactor}{4}
\providecommand{\BIBentryALTinterwordspacing}{\spaceskip=\fontdimen2\font plus
\BIBentryALTinterwordstretchfactor\fontdimen3\font minus
  \fontdimen4\font\relax}
\providecommand{\BIBforeignlanguage}[2]{{%
\expandafter\ifx\csname l@#1\endcsname\relax
\typeout{** WARNING: IEEEtran.bst: No hyphenation pattern has been}%
\typeout{** loaded for the language `#1'. Using the pattern for}%
\typeout{** the default language instead.}%
\else
\language=\csname l@#1\endcsname
\fi
#2}}
\providecommand{\BIBdecl}{\relax}
\BIBdecl

\bibitem{difonzo2007rumor}
N.~DiFonzo and P.~Bordia, \emph{Rumor psychology: Social and organizational
  approaches}.\hskip 1em plus 0.5em minus 0.4em\relax American Psychological
  Association Washington, DC, 2007, vol. 750.

\bibitem{castillo2011information}
C.~Castillo, M.~Mendoza, and B.~Poblete, ``Information credibility on
  twitter,'' in \emph{WWW}, 2011, pp. 675--684.

\bibitem{qazvinian2011rumor}
V.~Qazvinian, E.~Rosengren, D.~R. Radev, and Q.~Mei, ``Rumor has it:
  Identifying misinformation in microblogs,'' in \emph{EMNLP}, 2011, pp.
  1589--1599.

\bibitem{popat2017assessing}
K.~Popat, ``Assessing the credibility of claims on the web,'' in \emph{WWW '17
  Companion}, 2017, pp. 735--739.

\bibitem{yang2012automatic}
F.~Yang, Y.~Liu, X.~Yu, and M.~Yang, ``Automatic detection of rumor on sina
  weibo,'' in \emph{MDS '12}, 2012, p.~13.

\bibitem{jin2013epidemiological}
F.~Jin, E.~Dougherty, P.~Saraf, Y.~Cao, and N.~Ramakrishnan, ``Epidemiological
  modeling of news and rumors on twitter,'' in \emph{SNAKDD '13}.\hskip 1em
  plus 0.5em minus 0.4em\relax ACM, 2013, p.~8.

\bibitem{sampson2016leveraging}
J.~Sampson, F.~Morstatter, L.~Wu, and H.~Liu, ``Leveraging the implicit
  structure within social media for emergent rumor detection,'' in
  \emph{CIKM}.\hskip 1em plus 0.5em minus 0.4em\relax ACM, 2016, pp.
  2377--2382.

\bibitem{ma2017detect}
J.~Ma, W.~Gao, and K.-F. Wong, ``Detect rumors in microblog posts using
  propagation structure via kernel learning,'' in \emph{ACL}, 2017, pp.
  708--717.

\bibitem{ma2016detecting}
J.~Ma, W.~Gao, P.~Mitra, S.~Kwon, B.~J. Jansen, K.-F. Wong, and M.~Cha,
  ``Detecting rumors from microblogs with recurrent neural networks.'' in
  \emph{IJCAI}, 2016, pp. 3818--3824.

\bibitem{ma2018rumor}
J.~Ma, W.~Gao, and K.-F. Wong, ``Rumor detection on twitter with
  tree-structured recursive neural networks,'' in \emph{ACL (Volume 1: Long
  Papers)}, 2018, pp. 1980--1989.

\bibitem{yuan2019jointly}
C.~Yuan, Q.~Ma, W.~Zhou, J.~Han, and S.~Hu, ``Jointly embedding the local and
  global relations of heterogeneous graph for rumor detection,'' \emph{arXiv
  preprint arXiv:1909.04465}, 2019.

\bibitem{yao2019graph}
L.~Yao, C.~Mao, and Y.~Luo, ``Graph convolutional networks for text
  classification,'' in \emph{AAAI}, vol.~33, 2019, pp. 7370--7377.

\bibitem{zhang2019dual}
X.~Zhang, T.~Zhang, W.~Zhao, Z.~Cui, and J.~Yang, ``Dual-attention graph
  convolutional network,'' \emph{arXiv preprint arXiv:1911.12486}, 2019.

\bibitem{shi2019fine}
X.~Shi, L.~Xu, and P.~Wang, ``Fine-grained image classification combined with
  label description,'' in \emph{2019 IEEE 31st International Conference on
  Tools with Artificial Intelligence (ICTAI)}.\hskip 1em plus 0.5em minus
  0.4em\relax IEEE, 2019, pp. 1057--1064.

\bibitem{velivckovic2017graph}
P.~Veli{\v{c}}kovi{\'c}, G.~Cucurull, A.~Casanova, A.~Romero, P.~Lio, and
  Y.~Bengio, ``Graph attention networks,'' \emph{arXiv preprint
  arXiv:1710.10903}, 2017.

\bibitem{vaswani2017attention}
A.~Vaswani, N.~Shazeer, N.~Parmar, J.~Uszkoreit, L.~Jones, A.~N. Gomez,
  {\L}.~Kaiser, and I.~Polosukhin, ``Attention is all you need,'' in
  \emph{NeurIPS}, 2017, pp. 5998--6008.

\bibitem{xu2015empirical}
B.~Xu, N.~Wang, T.~Chen, and M.~Li, ``Empirical evaluation of rectified
  activations in convolutional network,'' \emph{arXiv preprint
  arXiv:1505.00853}, 2015.

\bibitem{zubiaga2016analysing}
A.~Zubiaga, M.~Liakata, R.~Procter, G.~W.~S. Hoi, and P.~Tolmie, ``Analysing
  how people orient to and spread rumours in social media by looking at
  conversational threads,'' \emph{PloS one}, vol.~11, no.~3, 2016.

\bibitem{liu2018early}
Y.~Liu and Y.-F.~B. Wu, ``Early detection of fake news on social media through
  propagation path classification with recurrent and convolutional networks,''
  in \emph{AAAI}, 2018.

\bibitem{zhao2015enquiring}
Z.~Zhao, P.~Resnick, and Q.~Mei, ``Enquiring minds: Early detection of rumors
  in social media from enquiry posts,'' in \emph{WWW}, 2015, pp. 1395--1405.

\bibitem{kwon2013prominent}
S.~Kwon, M.~Cha, K.~Jung, W.~Chen, and Y.~Wang, ``Prominent features of rumor
  propagation in online social media,'' in \emph{ICDM}, 2013, pp. 1103--1108.

\bibitem{ma2015detect}
J.~Ma, W.~Gao, Z.~Wei, Y.~Lu, and K.-F. Wong, ``Detect rumors using time series
  of social context information on microblogging websites,'' in \emph{CIKM},
  2015, pp. 1751--1754.

\bibitem{wu2015false}
K.~Wu, S.~Yang, and K.~Q. Zhu, ``False rumors detection on sina weibo by
  propagation structures,'' in \emph{ICDE}, 2015, pp. 651--662.

\bibitem{kingma2014adam}
D.~P. Kingma and J.~Ba, ``Adam: A method for stochastic optimization,''
  \emph{arXiv preprint arXiv:1412.6980}, 2014.

\bibitem{ruchansky2017csi}
N.~Ruchansky, S.~Seo, and Y.~Liu, ``Csi: A hybrid deep model for fake news
  detection,'' in \emph{CIKM}.\hskip 1em plus 0.5em minus 0.4em\relax ACM,
  2017, pp. 797--806.

\bibitem{huang2019deep}
Q.~Huang, C.~Zhou, J.~Wu, M.~Wang, and B.~Wang, ``Deep structure learning for
  rumor detection on twitter,'' in \emph{IJCNN}.\hskip 1em plus 0.5em minus
  0.4em\relax IEEE, 2019, pp. 1--8.

\end{thebibliography}
\end{document}